\documentclass[10pt]{article}

\newlength{\minitwocolumn}
\font\teneufm=eufm10
\font\seveneufm=eufm7
\font\fiveeufm=eufm5
\newfam\eufmfam
\textfont\eufmfam=\teneufm
\scriptfont\eufmfam=\seveneufm
\scriptscriptfont\eufmfam=\fiveeufm

\makeatletter
\@addtoreset{equation}{section}
\makeatother

\title{\bf
\Large{\bf 
Vertex operator approach to
semi-infinite spin chain : \\
recent progress}}
\begin{document}

\maketitle

\begin{center}
{Takeo Kojima}

{\it
Department of Mathematics and Physics,
Faculty of Engineering,
Yamagata University,\\
 Jonan 4-3-16, Yonezawa 992-8510, JAPAN\\
kojima@yz.yamagata-u.ac.jp}
\end{center}

~\\

\begin{abstract}
Vertex operator approach is a powerful method to study exactly solvable models.
We review recent progress of vertex operator approach to semi-infinite 
spin chain.
(1) The first progress is a generalization of boundary condition.
We study $U_q(\widehat{sl}(2))$ spin chain with a triangular boundary, 
which gives a generalization of diagonal boundary \cite{BB, BK}.
We give a bosonization of the boundary vacuum state. 
As an application, we derive a summation formulae of boundary magnetization.
(2) The second progress is a generalization of hidden symmetry.
We study supersymmetry $U_q(\widehat{sl}(M|N))$ spin chain 
with a diagonal boundary \cite{Kojima3}.
By now we have studied spin chain with a boundary,
associated with symmetry $U_q(\widehat{sl}(N))$, $U_q(A_2^{(2)})$ and
$U_{q,p}(\widehat{sl}(N))$ \cite{FK, Kojima1, Kojima2},
where bosonizations of vertex operators are realized by "monomial" .
However the vertex operator for $U_q(\widehat{sl}(M|N))$ 
is realized by "sum", a bosonization of boundary vacuum state 
is realized by "monomial".
\end{abstract}

\newpage

\section{Introduction}

There have been many developments in exactly solvable lattice models.
Various models were found to be solvable and various methods were invented to 
solve these models.
Vertex operator approach is a powerful method to study exactly solvable lattice models.
Solvability of lattice models is understood by means of 
commuting transfer matrix.
The half transfer matrices are called "vertex operators"
and are identified with the intertwiners of the 
irreducible highest weight representations of the quantum affine algebras
$U_q({g})$.
This identification is basis of vertex operator approach.
Vertex operator approach to boundary problem has been extended as 
generalizations of the theory on 
half-infinite $XXZ$ spin chain with a diagonal boundary
\cite{JKKKM}.
In this paper we review 
recent progress of vertex operator approach to semi-infinite spin chain
with a boundary.
We start from solutions of the boundary Yang-Baxter equation,
and introduce the transfer matrices in terms
of a product of vertex operators.
We diagonalize the transfer matrices by using bosonizations 
of the vertex operators, and study correlation functions.

The plan of the paper is as follows.
In section 2
we study $U_q(\widehat{sl}(2))$ spin chain with a triangular boundary, 
which is a generalization of diagonal boundary.
We give a bosonization of the boundary vacuum state, 
and calculate boundary magnetization.
In section 3
we study supersymmetry $U_q(\widehat{sl}(M+1|N+1))$ spin chain 
with a diagonal boundary.
We give bosonizations of boundary vacuum states.
In section 4 we summarize a conclusion.
Throughout this paper we use the following abbreviations.
\begin{eqnarray}
~[n]_q=\frac{q^n-q^{-n}}{q-q^{-1}},~~~
(z;p)_\infty=\prod_{m=0}^\infty (1-p^mz),~~~
\theta_m=\left\{\begin{array}{cc}
1& (m : {\rm even}),\\
0& (m : {\rm odd}).
\end{array}
\right.
\end{eqnarray}

\section{XXZ spin chain with a triangular boundary}

\subsection{Transfer matrix}

The first progress is a generalization of boundary condition.
We study
XXZ spin chain with a triangular boundary \cite{BB, BK}.
The Hamiltonian $H_B^{(\pm)}$ is given by
\begin{eqnarray}
H_B^{(\pm)}=-\frac{1}{2}
\sum_{k=1}^\infty (\sigma_{k+1}^x \sigma_k^x+\sigma_{k+1}^y \sigma_k^y+\Delta \sigma_{k+1}^z \sigma_k^z)
-\frac{1-q^2}{4q}\frac{1+r}{1-r}\sigma_1^z-\frac{s}{1-r} \sigma_1^\pm,
\label{def:Hamiltonian}
\end{eqnarray}
where $\sigma^x, \sigma^y, \sigma^z, \sigma^\pm$ are the standard 
Pauli matrices.
In what follows we set $V={\bf C}v_+ \oplus {\bf C} v_-$.
Consider the infinite dimensional vector space  $\cdots \otimes V_3 \otimes V_2 \otimes V_1$, where the matrices $V_j$ are copies of $V$.
Let us introduce the subspace ${\cal H}^{(i)}~(i=0,1)$
of the half-infinite spin chain by
\begin{eqnarray}
{\cal H}^{(i)}=Span \{
\cdots \otimes v_{p(N)} \otimes \cdots \otimes v_{p(2)} \otimes v_{p(1)}|~p(N)=(-1)^{N+i}~(N \gg 1)\},
\end{eqnarray}
where $p:{\bf N} \to \{\pm \}$.
Here we consider the model in the limit of 
half-infinite spin chain, in the massive regime where
$\Delta=\frac{q+q^{-1}}{2}$, $-1<q<0$, $-1\leq r \leq 1$, $s \in {\bf R}$.
The Hamiltonian $H_B^{(\pm)}$ acts on the subspace ${\cal H}^{(i)}$.
In Sklyanin's framework \cite{Sklyanin},  the transfer matrix 
$\widehat{T}_B^{(\pm,i)}(\zeta;r,s)$ that is a generating function of the Hamiltonian $H_B^{(\pm)}$ was 
introduced. It is built from two objects: the  $R$-matrix and the $K-$matrix. 
We introduces the $R$-matrix $R(\zeta)$ by
\begin{eqnarray}
R(\zeta)=\frac{1}{\kappa(\zeta)}
\left(
\begin{array}{cccc}
1& & & \\
 &\frac{\displaystyle (1-\zeta^2)q}{\displaystyle 1-q^2\zeta^2}&\frac{
\displaystyle
(1-q^2)\zeta}{
\displaystyle
1-q^2\zeta^2}& \\
 &\frac{
\displaystyle
(1-q^2)\zeta}{
\displaystyle
1-q^2\zeta^2}&\frac{
\displaystyle
(1-\zeta^2)q}{
\displaystyle
1-q^2\zeta^2}& \\
 & & &1
\end{array}
\right).
\end{eqnarray}
Here we have set
$\kappa(\zeta)=\zeta \frac{(q^4\zeta^2;q^4)_\infty (q^2/\zeta^2;q^4)_\infty}{
(q^4/\zeta^2;q^4)_\infty (q^2\zeta^2;q^4)_\infty}$.
The matrix elements of $R(\zeta) \in {\rm End}(V \otimes V)$ 
are given by
$R(\zeta)v_{\epsilon_1}\otimes v_{\epsilon_2}=
\sum_{\epsilon_1', \epsilon_2'=\pm} 
v_{\epsilon_1'}\otimes v_{\epsilon_2'}
R(\zeta)_{\epsilon_1' \epsilon_2'}^{\epsilon_1 \epsilon_2}$,
where the ordering of the index is given by
$v_+\otimes v_+, v_+\otimes v_-, v_-\otimes v_+, v_-\otimes v_-$.
$R_{i j}(\zeta)$ acts as
$R(\zeta)$ on the $i$-th and $j$-th components and as identity elsewhere.
The $R$-matrix $R(\zeta)$ satisfies the Yang-Baxter equation.
\begin{eqnarray}
R_{1 2}(\zeta_1/\zeta_2)
R_{1 3}(\zeta_1/\zeta_3)
R_{2 3}(\zeta_2/\zeta_3)
=
R_{2 3}(\zeta_2/\zeta_3)
R_{1 3}(\zeta_1/\zeta_3)
R_{1 2}(\zeta_1/\zeta_2).
\end{eqnarray}
The normalization factor $\kappa(\zeta)$
is determined by the following unitarity and crossing symmetry conditions:
$R_{12}(\zeta)R_{21}(\zeta^{-1})=1,~~~
R(\zeta)_{\epsilon_2 \epsilon_1'}^{\epsilon_2' \epsilon_1}=
R(-q^{-1}\zeta^{-1})_{-\epsilon_1 \epsilon_2}^{-\epsilon_1' \epsilon_2'}$.
Also, we introduce the triangular $K$-matrix 
$K^{(\pm)}(\zeta)=K^{(\pm)}(\zeta;r,s)$ 
by
\begin{eqnarray}
K^{(+)}(\zeta;r,s)
&=&
\frac{\varphi(\zeta^2;r)}{\varphi(\zeta^{-2};r)}
\left(\begin{array}{cc}
\frac{\displaystyle 1-r\zeta^2}{\displaystyle \zeta^2-r}&
\frac{\displaystyle s \zeta(\zeta^2-\zeta^{-2})}{
\displaystyle \zeta^2-r}\\
0&1
\end{array}
\right),
\label{def:K+}\\
K^{(-)}(\zeta;r,s)
&=&
\frac{\varphi(\zeta^2;r)}{\varphi(\zeta^{-2};r)}
\left(\begin{array}{cc}
\frac{\displaystyle
1-r\zeta^2}{
\displaystyle
\zeta^2-r}&
0\\
\frac{
\displaystyle
s \zeta (\zeta^2-\zeta^{-2})}{
\displaystyle
\zeta^2-r}&1
\end{array}
\right),
\label{def:K-}
\end{eqnarray}
where we have set
$\varphi(z;r)=
\frac{(q^4rz;q^4)_\infty 
(q^6z^2;q^8)_\infty}{
(q^2 rz;q^4)_\infty 
(q^8z^2;q^8)_\infty}$.
The matrix elements of $K^{(\pm)}(\zeta) \in {\rm End}(V)$ 
are given by
$K^{(\pm)}(\zeta)v_{\epsilon}=
\sum_{\epsilon'=\pm} 
v_{\epsilon'}
K^{(\pm)}(\zeta)_{\epsilon'}^{\epsilon}$,
where the ordering of the index is given by $v_+, v_-$.
The $K$-matrix $K^{(\pm)}(\zeta)$ satisfies 
the boundary Yang-Baxter equation :
\begin{eqnarray}
K_2^{(\pm)}(\zeta_2)
R_{21}(\zeta_1\zeta_2)
K_1^{(\pm)}(\zeta_1)
R_{12}(\zeta_1/\zeta_2)
=R_{21}(\zeta_1/\zeta_2)
K_1^{(\pm)}(\zeta_1)
R_{12}(\zeta_1\zeta_2)K_2^{(\pm)}(\zeta_2).
\label{eqn:BYBE}
\end{eqnarray}
The normalization factor $\varphi(z;r)$
is determined by the following boundary unitarity and 
boundary crossing symmetry :
$K^{(\pm)}(\zeta)K^{(\pm)}(\zeta^{-1})=1$, 
${K^{(\pm)}}(-q^{-1}\zeta^{-1})_{\epsilon_1}^{\epsilon_2}=
\sum_{\epsilon_1', \epsilon_2'=\pm}
R(-q\zeta^2)_{\epsilon_1' -\epsilon_2'}^{-\epsilon_1 \epsilon_2}
{K^{(\pm)}}(\zeta)_{\epsilon_2'}^{\epsilon_1'}$.
We introduce 
the vertex operators $\widehat{\Phi}_\epsilon^{(1-i,i)}(\zeta)$
$(\epsilon=\pm)$
which act  on the space ${\cal H}^{(i)}$ $(i=0,1)$.
Matrix elements are given by products of the $R$-matrix as follows:
\begin{eqnarray}
(\widehat{\Phi}_\epsilon^{(1-i,i)}(\zeta))
^{\cdots p(N)' \cdots p(2)' p(1)'}_{
\cdots p(N) \cdots p(2)~p(1)}
&=&
\lim_{N \to \infty}
\sum_{\mu(1), \mu(2), \cdots, \mu(N)=\pm}\prod_{j=1}^N R(\zeta)_{\mu(j-1)~p(j)}^{\mu(j)~p(j)'},
\end{eqnarray}
where $\mu(0)=\epsilon$ and $\mu(N)=(-1)^{N+1-i}$.
We expect that 
the vertex operators $\widehat{\Phi}_\epsilon^{(1-i,i)}(\zeta)$ 
give rise to well-defined operators.
We set
$\widehat{\Phi}_{\epsilon}^{*(1-i,i)}(\zeta)=
\widehat{\Phi}_{-\epsilon}^{(1-i,i)}(-q^{-1}\zeta)$.
Following the strategy \cite{JKKKM} we introduce
the transfer matrix $\widehat{T}_B^{(\pm,i)}(\zeta; r,s)$ using the 
vertex operators.
\begin{eqnarray}
\widehat{T}_B^{(\pm,i)}(\zeta;r,s)=\sum_{\epsilon_1, \epsilon_2=\pm}
\widehat{\Phi}_{\epsilon_1}^{* (i,1-i)}(\zeta^{-1})K^{(\pm)}(\zeta;r,s)_{\epsilon_1}^{\epsilon_2}
\widehat{\Phi}_{\epsilon_2}^{(1-i,i)}(\zeta).
\label{def:phys-transfer}
\end{eqnarray}
Heuristic arguments suggest that the transfer matrix commutes :
\begin{eqnarray}
~[\widehat{T}_B^{(\pm,i)}(\zeta_1;r,s),
\widehat{T}_B^{(\pm,i)}(\zeta_2;r,s)]=0~~~~{\rm for~any}~~\zeta_1, \zeta_2.
\end{eqnarray}
The Hamiltonian $H_B^{(\pm)}$ (\ref{def:Hamiltonian}) is obtained as
\begin{eqnarray}
\left.\frac{d}{d\zeta}\widehat{T}_B^{(\pm,i)}(\zeta;r,s)\right|_{\zeta=1}=\frac{4q}{1-q^2}H_B^{(\pm)}+{\rm const}.
\end{eqnarray}
We are interested in diagonalization of the transfer matrix 
$\widehat{T}_B^{(\pm,i)}(\zeta;r,s)$.

\subsection{Vertex operator approach}

We formulate the vertex operator approach to 
the half-infinite $XXZ$ spin chain with a triangular boundary.
Let $V_\zeta$ the
evaluation representation of $U_q(\widehat{sl}(2))$.
Let $V(\Lambda_i)$
the irreducible highest weight $U_q(\widehat{sl}(2))$ representation
with the fundamental weights $\Lambda_i$ $(i=0,1)$.
We introduce the vertex operators $\Phi^{(1-i,i)}_\epsilon(\zeta)$
as the intertwiner of $U_q(\widehat{sl}(2))$:
\begin{eqnarray}
\Phi^{(1-i,i)}(\zeta) : V(\Lambda_i) \longrightarrow V(\Lambda_{1-i}) \otimes V_\zeta,
&&
\Phi^{(1-i,i)}(\zeta) \cdot x=\Delta(x) \cdot \Phi^{(1-i,i)}(\zeta),
\end{eqnarray}
for $x \in U_q(\widehat{sl}(2))$.
We set the elements of the vertex operators :
$\Phi^{(1-i,i)}(\zeta)=\sum_\epsilon \Phi_\epsilon^{(1-i,i)}(\zeta) \otimes v_\epsilon$.
We set
${\Phi}_{\epsilon}^{*(1-i,i)}(\zeta)=
{\Phi}_{-\epsilon}^{(1-i,i)}(-q^{-1}\zeta)$.
Following the strategy of \cite{JKKKM}, 
as the generating function of the Hamiltonian $H_B^{(\pm)}$ we introduce
the ``renormalized" transfer matrix ${T}_B^{(\pm,i)}(\zeta; r,s)$ :
\begin{eqnarray}
{T}_B^{(\pm,i)}(\zeta;r,s)=g \sum_{\epsilon_1, \epsilon_2=\pm}
{\Phi}_{\epsilon_1}^{* (i,1-i)}(\zeta^{-1})K^{(\pm)}(\zeta;r,s)_{\epsilon_1}^{\epsilon_2}
{\Phi}_{\epsilon_2}^{(1-i,i)}(\zeta),~
g=\frac{(q^2;q^4)_\infty}{(q^4;q^4)_\infty}.
\label{def:math-transfer}
\end{eqnarray}
Following strategy \cite{JKKKM}, 
we study our problem upon the following identification:
\begin{eqnarray}
T_B^{(\pm,i)}(\zeta;r,s)=\widehat{T}_B^{(\pm,i)}(\zeta;r,s),
\Phi_\epsilon^{(1-i,i)}(\zeta)=
\widehat{\Phi}_\epsilon^{(1-i,i)}(\zeta),
\Phi_\epsilon^{*(1-i,i)}(\zeta)=
\widehat{\Phi}_\epsilon^{*(1-i,i)}(\zeta).
\end{eqnarray}
The point of using the 
vertex operators 
$\Phi_\epsilon^{(1-i,i)}(\zeta)$
associated with $U_q(\widehat{sl}(2))$ is that they are well-defined objects, 
free from the difficulty of divergence. 
It is convenient to diagonalize
the ``renormalized" transfer matrix ${T}_B^{(\pm,i)}(\zeta; r,s)$
instead of the Hamiltonian $H_B^{(\pm)}$.

\subsection{Boundary vacuum state}

We are interested in bosonizations of
the boundary vacuum states 
$~_B\langle i;\pm|$ given by
\begin{eqnarray}
~_B\langle i; \pm|T_B^{(\pm,i)}(\zeta;r,0)=
\Lambda^{(i)}(\zeta;r)_B\langle i; \pm|,
\end{eqnarray}
for $i=0,1$.
Here we have set $\Lambda^{(0)}(\zeta;r)=1$ and
$\Lambda^{(1)}(\zeta;r)=\frac{1}{\zeta^2}\frac{\Theta_{q^4}(r\zeta^2)
\Theta_{q^4}(q^2r\zeta^{-2})}{
\Theta_{q^4}(r\zeta^{-2})\Theta_{q^4}(q^2r\zeta^2)}$,
where $\Theta_p(z)=(p;p)_\infty (z;p)_\infty (p/z;p)_\infty$.
We introduce bosons
$a_m~(m\neq 0)$ and
the zero-mode operator $\partial, \alpha$ by
\begin{eqnarray}
[a_m,a_n]=\delta_{m+n,0}\frac{[2m]_q[m]_q}{m}~~~(m,n\neq 0),~~~
[\partial,\alpha]=2.
\end{eqnarray}
The relation between the zero-mode and the fundamental weights
are given by
$[\partial,\Lambda_0]=0$ and $\Lambda_1=\Lambda_0+\frac{\alpha}{2}$.
Using the bosonization of the vertex operators 
$\Phi_\epsilon^{(1-i,i)}(\zeta)$ we have
a bosonization of the boundary vacuum state. 
The boundary vacuum states $~_B\langle i ;\pm|$
are realized by
\begin{eqnarray}
&&
~_B\langle 0; +|=~_B\langle 0|\exp_q\left(-s f_0\right),
~~~_B\langle 1;+|=~_B\langle 1|\exp_{q^{-1}}\left(-\frac{~s~}{r q} e_1 q^{-h_1}\right),
\\
&&~_B\langle 0; -|=~_B\langle 0|\exp_{q^{-1}}\left(\frac{~s~}{q} e_0 q^{-h_0}\right),
~~~_B\langle 1;-|=~_B\langle 1|\exp_{q}\left(\frac{~s~}{r} f_1\right),
\end{eqnarray}
where we have used $q$-exponential $\exp_q(x)=\sum_{n=0}^\infty
\frac{q^{\frac{n(n-1)}{2}}}{[n]_q!}x^n$.
Here $~_B\langle i|$ are given by
\begin{eqnarray}
~_B\langle i|=\langle i|\exp\left(G_i\right), && G_i=-
\frac{1}{2}\sum_{n=1}^\infty \frac{n q^{-2n}}{[2n]_q[n]_q} a_n^2
+\sum_{n=1}^\infty \delta_n^{(i)}a_n,~~~
\langle i|=1 \otimes e^{-\Lambda_i}.
\label{def:vacdual}
\end{eqnarray}
where we have set 
\begin{eqnarray}
\delta_n^{(i)}&=&
\theta_n \frac{q^{-{3n}/{2}}(1-q^n)}{[2n]_q}+\left\{
\begin{array}{cc}
-\frac{\displaystyle q^{-{5n}/{2}}r^n}{\displaystyle [2n]_q}&~(i=0),\\
+\frac{\displaystyle q^{-{n}/{2}}r^{-n}}{\displaystyle [2n]_q}&~(i=1).
\end{array}
\right.\label{def:delta}
\end{eqnarray}
The boundary vacuum states $|\pm;i\rangle_B$ are realized similarly.

\subsection{Boundary magnetization}

In this section we study the boundary magnetization.
Let ${\cal E}_{\epsilon, \epsilon'}$ be 
the matrix $E_{\epsilon, \epsilon'}$
at the first site of the space ${\cal H}^{(i)}$.
We have a realization of this local operator
\begin{eqnarray}
{\cal E}_{\epsilon, \epsilon'}=g \left.
\Phi_{\epsilon}^{*(i,1-i)}(-q^{-1}\zeta)
\Phi_{\epsilon'}^{(1-i,i)}(\zeta)\right|_{\zeta=1},~~~
g=\frac{(q^2;q^4)_\infty}{(q^4;q^4)_\infty}.
\end{eqnarray}
Hence, using the bosonizations of the vertex operators,
the Chevalley generators $e_j, f_j, h_j$ $(j=0,1)$, 
and the boundary vacuum states,
we calculate the following vacuum expectation values.
\begin{eqnarray}
\frac{~_B\langle i; \pm|{\cal E}_{\epsilon,\epsilon'}|\pm; i \rangle_B}
{
~_B\langle i; \pm|\pm; i \rangle_B}.
\end{eqnarray}
For instance,
the boundary magnetizations are derived:
\begin{eqnarray}
\frac{~_B\langle 0;-|\sigma_1^z |-;0\rangle_B}{~_B\langle 0;-|-
;0\rangle_B}
&=&-1-2(1-r)^2 \sum_{n=1}^\infty
\frac{(-q^2)^n}{(1-rq^{2n})^2},\\
\frac{~_B\langle 0;-|\sigma_1^+ |-;0\rangle_B}{~_B\langle 0;-|-;0\rangle_B}
&=&s
\left(2+(1-r)\sum_{n=1}^\infty
(-q^2)^n\frac{2q^{2n}-r(1+q^{4n})}{(1-rq^{2n})^2}
\right),\\
\frac{~_B\langle 0;-|\sigma_1^- |-;0\rangle_B}{~_B\langle 0;-|-;0\rangle_B}
&=&0.
\end{eqnarray}
This is {\bf main result} of the paper \cite{BK}.
%We have integral representations of
%general correlation functions \cite{BK}.

\section{$U_q(\widehat{sl}(M+1|N+1))$ spin chain with a diagonal boundary}

\subsection{Transfer matrix}

The second progress is a generalization of hidden symmetry.
We study
$U_q(\widehat{sl}(M+1|N+1))$ spin chain with a diagonal boundary
\cite{Kojima3}.
Let us set $-1<q<0$ and $r \in {\bf R}$.
Let us set $M,N=0,1,2,\cdots (M \neq N)$ and $L,K=1,2,\cdots,M+N+2$.
For simplicity we assume the condition $L+K\leq M+1$. 
(More general cases are studied in \cite{Kojima3}.)
Let us introduce the signatures $\nu_i$ $(i=1,2,\cdots,M+N+2)$ by 
$\nu_1=\cdots=\nu_{M+1}=+$, $\nu_{M+2}=\cdots=\nu_{M+N+2}=-$.
Let us set the vector spaces 
$V_1=\oplus_{j=1}^{M+1}{\bf C}v_j$ and 
$V_0=\oplus_{j=1}^{N+1}{\bf C}v_{M+1+j}$.
In this section we set $V=V_1 \oplus V_0$.
The ${\bf Z}_2$-grading of the basis $\{v_j\}_{1\leq j \leq M+N+2}$ of $V$ is chosen to be 
$\left[v_j\right]=\frac{\nu_j+1}{2}$ $(j=1, 2, \cdots, M+N+2)$.
A linear operator $S \in {\rm End}(V)$ is represented in the form of
a $(M+N+2)\times(M+N+2)$ matrix : $S v_j=\sum_{i=1}^{M+N+2}v_i S_{i,j}$.
The ${\bf Z}_2$-grading of $(M+N+2)\times(M+N+2)$ matrix 
$(S_{i,j})_{1\leq i,j \leq M+N+2}$ is defined by
$[S]=[v_i]+[v_j]~(mod.2)$ if RHS of the equation 
does not depend on $i$ and $j$ such that $S_{i,j}\neq 0$.
We define the action of the operator
$S_1 \otimes \cdots \otimes S_n$ where $S_j \in {\rm End}(V)$
have ${\bf Z}_2$-grading.
\begin{eqnarray}
&&
S_1 \otimes S_2 \otimes \cdots \otimes S_n \cdot
v_{j_1}\otimes v_{j_2} \otimes \cdots \otimes v_{j_n}\nonumber\\
&=&
\exp\left(\pi \sqrt{-1}~
\sum_{k=1}^n [S_k] \sum_{l=1}^{k-1}[v_{j_l}]\right) 
S_1 v_{j_1}\otimes S_2 v_{j_2} \otimes \cdots \otimes 
S_n v_{j_n}.
\end{eqnarray}
We set the $R$-matrix 
$R(z) \in {\rm End}(V \otimes V)$ for
$U_q(\widehat{sl}(M+1|N+1))$
as follows.
\begin{eqnarray}
R(z)=r(z)\bar{R}(z),~~~
\bar{R}(z)v_{j_1}\otimes v_{j_2}=\sum_{k_1,k_2=1}^{M+N+2}
v_{k_1}\otimes v_{k_2}
\bar{R}(z)_{k_1,k_2}^{j_1,j_2}.
\label{def:R-matrix1}
\end{eqnarray}
Here we have set
\begin{eqnarray}
\bar{R}(z)_{j,j}^{j,j}
&=&
\left\{
\begin{array}{cc}
-1& (1\leq j \leq M+1),\\
-\frac{\displaystyle
(q^2-z)}{\displaystyle
(1-q^2 z)}& (M+2\leq j \leq M+N+2),
\end{array}
\right.
\label{def:R-matrix2}
\\
\bar{R}(z)_{i,j}^{i,j}&=&\frac{(1-z)q}{(1-q^2 z)}
~~~(1\leq i \neq j \leq M+N+2),
\label{def:R-matrix3}
\\
\bar{R}(z)_{i,j}^{j,i}&=&
\left\{
\begin{array}{cc}
(-1)^{ [v_i] [v_j] }\frac{\displaystyle
(1-q^2)}{
\displaystyle
(1-q^2 z)}&
(1\leq i < j \leq M+N+2),
\label{def:R-matrix4}
\\
(-1)^{ [v_i] [v_j] }
\frac{
\displaystyle
(1-q^2)z}{
\displaystyle
(1-q^2 z)}&
(1\leq j < i \leq M+N+2),
\end{array}
\right.
\label{def:R-matrix5}\\
\bar{R}(z)_{i,j}^{i,j}&=&0~~~~~{\rm otherwise}.
\end{eqnarray}
Here we have set
\begin{eqnarray}
r(z)=z^{\frac{1-M+N}{M-N}}\exp\left(-\sum_{m=1}^\infty
\frac{[(M-N-1)m]_q}{m[(M-N)m]_q}q^m(z^m-z^{-m})\right).
\label{def:r}
\end{eqnarray}
The $R$-matrix
${R}(z)$
satisfies the graded Yang-Baxter equation.
\begin{eqnarray}
{R}_{1 2}(z_1/z_2)
{R}_{1 3}(z_1/z_3)
{R}_{2 3}(z_2/z_3)=
{R}_{2 3}(z_2/z_3)
{R}_{1 3}(z_1/z_3)
{R}_{1 2}(z_1/z_2).
\label{def:gYBE}
\end{eqnarray}
We set the diagonal $K$-matrix $K(z) \in {\rm End}(V)$ 
for $U_q(\widehat{sl}(M+1|N+1))$ as follows. 
\begin{eqnarray}
&&K(z)=z^{-\frac{2M}{M-N}}
\frac{\varphi(z)}{
\varphi(z^{-1})}
\bar{K}(z),~~~\bar{K}(z)v_j=\sum_{k=1}^{M+N+2}
v_k \delta_{j,k} \bar{K}(z)_j^j,
\end{eqnarray}
where we have set
\begin{eqnarray}
\bar{K}(z)_j^j=
\left\{\begin{array}{cc}
1& (1\leq j \leq L),\\
\frac{\displaystyle 1-r/z}{\displaystyle 1-r z}&
(L+1 \leq j \leq L+K),\\
z^{-2}& (L+K+1 \leq j \leq M+N+2).
\end{array}
\right.\label{def:tildeK3}
\end{eqnarray}
Here we have set
\begin{eqnarray}
\varphi(z)&=&
\exp\left(
\sum_{m=1}^\infty
\frac{[2(N+1)m]_q}{m [2(M-N)m]_q}z^{2m}+
\sum_{j=1}^M
\sum_{m=1}^\infty
\frac{[2(M-N-j)m]_q}{2m [2(M-N)m]_q}(1-q^{2m})z^{2m}
\right.\\
&+&\sum_{j=M+2}^{M+N+1}\sum_{m=1}^\infty
\frac{[2(-M-N-2-j)m]_q}{2m [2(M-N)m]_q}(1+q^{2m})z^{2m}
-\sum_{m=1}^\infty
\frac{[(M-N-1)m]_q}{2m[(M-N)m]_q}q^mz^{2m}
\nonumber\\
&+&\left.
\sum_{m=1}^\infty
\left\{\frac{[(-M+N+L)m]_q}{m [(N-M)m]_q} 
(rq^{-L}z)^m
+\frac{[(-M+N+L+K)m]_q}{m [(M-N)m]_q}(q^{L-K}z/r)^m \right\}
\right).
\nonumber
\end{eqnarray}
The $K$-matrix $K(z) \in {\rm End}(V)$
satisfies the graded boundary Yang-Baxter equation
\begin{eqnarray}
{K}_2(z_2)
{R}_{21}(z_1z_2)
{K}_1(z_1)
{R}_{12}(z_1/z_2)=
{R}_{21}(z_1/z_2)
{K}_1(z_1)
{R}_{12}(z_1z_2)
{K}_2(z_2).
\label{def:gBYBE}
\end{eqnarray}
We introduce 
the vertex operators $\widehat{\Phi}_j(z)$
and the dual vertex operators 
$\widehat{\Phi}_j^{*}(z)$ for $j=1,2,\cdots,M+N+2$.
Matrix elements are given by products of the $R$-matrix
\begin{eqnarray}
(\widehat{\Phi}_j(z))
^{\cdots p(N)' \cdots p(2)' p(1)'}_{
\cdots p(N) \cdots p(2)~p(1)}
&=&
\lim_{n \to \infty}
\sum_{\mu(1), \mu(2), \cdots, \mu(n)=1}^{M+N+2}
\prod_{j=1}^n R(z)_{\mu(j-1)~p(j)}^{\mu(j)~p(j)'},\\
(\widehat{\Phi}_j^*(z))
^{\cdots p(N)' \cdots p(2)' p(1)'}_{
\cdots p(N) \cdots p(2)~p(1)}
&=&
\lim_{n \to \infty}
\sum_{\mu(1), \mu(2), \cdots, \mu(n)=1}^{M+N+2}
\prod_{j=1}^n R(z)_{p(j)~\mu(j)}^{p(j)'\mu(j-1)},
\end{eqnarray}
where $\mu(0)=j$.
We expect that 
the vertex operators $\widehat{\Phi}_j(z)$ 
and $\widehat{\Phi}_j^*(z)$ 
give rise to well-defined operators.
Let us set the transfer matrix $\widehat{T}_B(z)$ by
\begin{eqnarray}
\widehat{T}_B(z)
=\sum_{j=1}^{M+N+2}
\widehat{\Phi}_j^*(z^{-1})K(z)_j^j \widehat{\Phi}_j(z)(-1)^{[v_j]}.
\end{eqnarray}
Heuristic arguments suggest that the transfer matrix commutes :
\begin{eqnarray}
~[\widehat{T}_B(z_1),\widehat{T}_B(z_2)]=0~~~{\rm for~any}~z_1,z_2.
\end{eqnarray}
The Hamiltonian of this model $H_B$ is given by
\begin{eqnarray}
H_B=\frac{d}{dz}
T_B(z)|_{z=1}=\sum_{j=1}^{\infty}h_{j,j+1}
+\frac{1}{2}\frac{d}{dz}K_1(z)|_{z=1},
\end{eqnarray}
where $h_{j,j+1}=P_{j,j+1}\frac{d}{dz}R_{j,j+1}(z)|_{z=1}$.

\subsection{Vertex operator approach}

We formulate the vertex operator approach to $U_q(\widehat{sl}(M+1|N+1))$
spin chain with a diagonal boundary \cite{Kojima3}.
Let $V_z$ the evaluation representation of $U_q(\widehat{sl}(M+1|N+1))$
and $V_z^{* S}$ its dual.
Let $L(\lambda)$ the irreducible highest representation with 
level-$1$ highest weight $\lambda$.
We introduce the vertex operators 
$\Phi(z)$ and $\Phi^*(z)$
as the intertwiners of $U_q(\widehat{sl}(M+1|N+1))$ :
\begin{eqnarray}
&&
\Phi(z) : L(\lambda)\rightarrow L(\mu)\otimes V_z,~~~
\Phi(z)\cdot x=\Delta(x)\cdot \Phi(z),\\
&&
\Phi^*(z) : L(\mu)\rightarrow L(\lambda)\otimes V_z^{*S},~~~
\Phi^*(z)\cdot x=\Delta(x)\cdot \Phi^*(z),
\end{eqnarray}
for $x \in U_q(\widehat{sl}(M+1|N+1))$.
We expand the vertex operators
$\Phi(z)=\sum_{j=1}^{M+N+2}
\Phi_j(z) \otimes v_j$, $\Phi^*(z)=\sum_{j=1}^{M+N+2}
\Phi_j^*(z) \otimes v_j^*$.
We set the "normalized" transfer matrix 
$T_B(z)$ by
\begin{eqnarray}
T_B(z)
=g \sum_{j=1}^{M+N+2}\Phi_j^*(z^{-1})
K^{(i)}(z)_j^j \Phi_j(z)(-1)^{[v_j]},
\label{dfn:infinite-transfer}
\end{eqnarray}
where we have used 
$g=e^{\frac{\pi \sqrt{-1} M}{2(M-N)}}
\exp\left(
-\sum_{m=1}^\infty
\frac{[(M-N-1)m]_q}{m[(M-N)m]_q}q^m
\right)$.
Following the strategy proposed in
\cite{JKKKM}, 
we consider our problem upon the following identification.
\begin{eqnarray}
T_B(z)=\widehat{T}_B(z),~~~
\Phi_j(z)=\widehat{\Phi}_j(z),~~~
\Phi_j^*(z)=\widehat{\Phi}_j^*(z).
\end{eqnarray}
The point of using the vertex operators 
$\Phi_j(z), \Phi_j^*(z)$
is that they are well-defined objects, 
free from the difficulty of divergence. 
It is convenient to diagonalize
the ``renormalized" transfer matrix ${T}_B(z)$
instead of the Hamiltonian $H_B$.

\subsection{Boundary vacuum state}

In this section we give a bosonization of the boundary vacuum state
$\langle B|$ given by
\begin{eqnarray}
\langle B|T_B(z)=\langle B|.\label{eqn:main}
\end{eqnarray}
Let us introduce the bosons and the zero-mode operator \cite{KSU} by
\begin{eqnarray}
a_n^k, b_n^l, c_n^l, Q_{a^k}, Q_{b^l}, Q_{c^l},
~~~(n \in {\bf Z}, k=1,2,\cdots,M+1, l=1,2,\cdots, N+1),
\end{eqnarray}
satisfying the following commutation relations.
\begin{eqnarray}
&&[a_m^i, a_n^j]=\delta_{i,j}\delta_{m+n,0}\frac{[m]_q^2}{m},
~[a_0^i, Q_{a^j}]=\delta_{i,j},~[a_0^i,a_0^j]=0,\\
&&[b_m^i,b_n^j]=-\delta_{i,j}\delta_{m+n,0}\frac{[m]_q^2}{m},
~[b_0^i,Q_{b^j}]=-\delta_{i,j},
~[b_0^i,b_0^j]=0,\\
&&[c_m^i,c_n^j]=\delta_{i,j}\delta_{m+n,0}\frac{[m]_q^2}{m},
~[c_0^i,Q_{c^j}]=\delta_{i,j},
~[c_0^i,c_0^j]=0.
\end{eqnarray}
Let us introduce the generating function 
$c^i(z)=-\sum_{n \neq 0}\frac{c^i_n}{[n]_q}z^{-n}
+Q_{c^i}+c_{0}^i {\rm log} z$.
We introduce the
projection operator $\xi_0=\prod_{j=1}^{N+1}\xi_0^j$ 
and $\eta_0=\prod_{j=1}^{N+1}\eta_0^j$, 
where we have set
$\xi^j(z)=\sum_{m \in {\bf Z}}\xi_m^j z^{-m}=:e^{-c^j(z)}:$,
$\eta^j(z)=\sum_{m \in {\bf Z}}\eta_m^j z^{-m-1}=:e^{c^j(z)}:$.
%For instance the bosonizations of the vertex operators 
%$\Phi^*_{M+1+j}(z)$ $(j=1,2,\cdots, N+1)$
%are written by the sum 
%$\sum_{\epsilon_1,\epsilon_2,\cdots,\epsilon_j}$ as follows 
%Here we skip the detailed explanations. See \cite{Kojima3}.
%\begin{eqnarray}
%\Phi_{M+1+j}^*(z)
%&=&
%\sum_{\epsilon_1,\epsilon_2,\cdots,\epsilon_j=\pm}
%e^{\frac{M \pi \sqrt{-1}}{M-N}}q^{j-1}(q-q^{-1})^M(qz)^{-1}
%\prod_{k=1}^j \epsilon_k \prod_{k=1}^{M+j}
%\int \frac{dw_k}{2\pi \sqrt{-1}w_k}
%\nonumber\\
%&\times&
%\frac{1}{
%\displaystyle
%\prod_{k=0}^M (1-qw_k/w_{k+1})(1-qw_{k+1}/w_k)
%\prod_{k=1}^{j-1}(1-q^{\epsilon_k}w_{M+k}/w_{M+k+1})}\\
%&\times&
%\eta_0 \xi_0 :\phi_1^*(z)
%X_1^{-}(qw_1)\cdots X_M^{-}(qw_M)
%X_{M+1,\epsilon_1}^{-}(qw_{M+1})\cdots X_{M+j,\epsilon_j}^{-}(qw_{M+j}):
%\eta_0 \xi_0.\nonumber
%\end{eqnarray}
Using the bosonizations of the vertex operators we have a bosonization of the boundary vacuum state $\langle B|$.
However the vertex operator for $U_q(\widehat{sl}(M+1|N+1))$ 
is realized by "sum", a bosonization of boundary vacuum state 
is realized by "monomial".
Let us set the highest weight vector 
$v_{\Lambda_{M+1}}^*=\langle 0
|e^{-\beta \sum_{i=1}^{M+1}Q_{a^i}
+(1-\beta)\sum_{j=1}^{N+1}Q_{b^j}
+\sum_{j=1}^{N+1}Q_{c^j}}$,
where $\langle 0|$ satisfying
$\langle 0|a_n^i=\langle 0|b_n^j=\langle 0|c_n^j=0$
for $n \geq 0$ and $1\leq i \leq M+1$, $1\leq j \leq N+1$.
Let us set
$h_{i,m}^{*}=\sum_{j=1}^{M+N+1}
\frac{[\alpha_{i,j}m]_q[\beta_{i,j}m]_q}{[(M-N)m]_q[m]_q}h_{j,m}$,
where we have used 
$h_{i,m}=a_m^i q^{-|m|/2}-a_{m}^{i+1} q^{|m|/2}$,
$h_{M+1,m}=a_m^{M+1}q^{-|m|/2}+b_m^1q^{-|m|/2}$,
and 
$h_{M+1+j,m}=-b_m^j q^{|m|/2}+b_m^{j+1}q^{-|m|/2}$.
Here we have set
\begin{eqnarray}
\alpha_{i,j}&=&\left\{\begin{array}{cc}
{\rm Min}(i,j)&~({\rm Min}(i,j) \leq M+1),\\
2(M+1)-{\rm Min}(i,j)&~({\rm Min}(i,j)>M+1),
\end{array}
\right.\\
\beta_{i,j}&=&\left\{\begin{array}{cc}
M-N-{\rm Max}(i,j)&~({\rm Max}(i,j)\leq M+1),\\
-M-N-2+{\rm Max}(i,j)&~({\rm Max}(i,j)>M+1).
\end{array}\right.
\end{eqnarray}
A bosonization of the boundary vacuum state $\langle B|$ is given by
\begin{eqnarray}
\langle B|=v_{\Lambda_{M+1}}^* \exp\left(G \right) \cdot \eta_0 \xi_0.
\end{eqnarray}
Here we have set
the bosonic operator $G$ by
\begin{eqnarray}
G&=&-\frac{1}{2}\sum_{j=1}^{M+N+1}
\sum_{m=1}^\infty
\frac{m q^{-2m}}{[m]_q^2}h_{j,m} h_{j,m}^{*}
-\frac{1}{2}\sum_{j=1}^{N+1}
\sum_{m=1}^\infty
\frac{m q^{-2m}}{[m]_q^2}c_m^j c_m^j\nonumber\\
&+&\sum_{j=1}^{M+N+1}
\sum_{m=1}^\infty
\beta_{j,m}^{(3)}h_{j,m}^{*}
+\sum_{j=1}^{N+1}
\sum_{m=1}^\infty
\gamma_{j,m} c_m^j,
\label{def:G}
\end{eqnarray}
where we have used
\begin{eqnarray}
\beta_{j,m}^{(3)}&=&\beta_{j,m}^{(1)}-\frac{\displaystyle
r^mq^{(-L-3/2) m}}{\displaystyle
[m]_q}\delta_{j,L}-\frac{
\displaystyle q^{(L-K-3/2) m}/r^m
}{
\displaystyle [m]_q
}\delta_{j,L+K},~
\gamma_{j,m}=-\frac{q^{-m}}{[m]_q}\theta_m,\\
\beta_{j,m}^{(1)}
&=&\left\{
\begin{array}{cc}
\frac{\displaystyle
q^{-3m/2}-q^{-m/2}}{
\displaystyle
[m]_q}\theta_m 
&(1\leq j \leq M),\\
\frac{\displaystyle
-2 q^{-3m/2}
}{\displaystyle
[m]_q}\theta_m &(j=M+1),\\
\frac{\displaystyle q^{-3m/2}+q^{-m/2}}{\displaystyle [m]_q}\theta_m
&
(M+2\leq j \leq M+N+1).
\end{array}
\right.
\end{eqnarray}
This is {\bf main result} of the paper \cite{Kojima3}.

\section{Conclusion}
From the above progress, we suppose that
the boundary vacuum state $\langle B|$
of semi-infinite 
$U_q({g})$
spin chain with a triangular boundary
is realized as follows.
\begin{eqnarray}
\langle B|=\langle vac|\exp\left({\cal B}\right)
\exp_q\left({\cal C}\right),
\end{eqnarray}
where ${\cal B}$ is a quadratic expression in the bosons and 
${\cal C}$ is a simple expression
in the Chevalley generators.
We would like to check this conjecture in the future.

\subsection*{Acknowledgements}
The authors would like to thank
Pascal Baseilhac for fruitful collaboration.
The author would like to thank 
Laboratoire de Math\'ematiques et Physique Th\'eorique, Universit\'e de Tours
for kind invitation and warm hospitality during his stay in March 2013.
This work is supported by the Grant-in-Aid for 
Scientific Research {\bf C} (21540228) from JSPS and Visiting professorship from CNRS.

\begin{appendix}

\end{appendix}

\end{document}